# THE INFORMATION CONTENT OF REDSHIFT AND VELOCITY SURVEYS

A. Yahil
*State University of New York, Stony Brook, NY, U.S.A.*

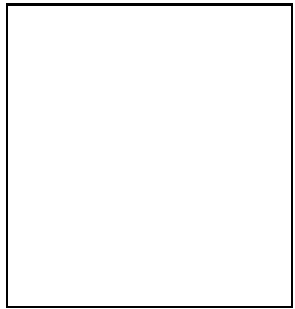

**Abstract**

The pixon-based image reconstruction of Puetter and Piña has achieved significant improvements over other methods in higher spatial resolution, greater sensitivity to faint sources, and immunity to the production of spurious artifacts and signal-correlated residuals. The same technique may be used for those problems of large-scale structure which allow for variable smoothing. The comparison of different datasets is not impaired by the variable smoothing, because a common underlying density/potential field, whatever its smoothing, can be applied to all datasets. By making optimal use of the combined datasets, the pixon method could therefore yield the most sensitive determination of the cosmological density parameter, $\Omega$, from redshift and velocity surveys.

## 1 Introduction

The mapping of large-scale structures and velocities shares fundamental similarities with image reconstruction/restoration.[1] In both cases, the problem consists of solving an equation which can be written symbolically as

$$D = H \otimes I + N \quad . \tag{1}$$

Here $D$ is the array of data points, $I$ is the underlying field to be determined, image or density/potential, and $N$ is the noise in the data. The operator $H$ transforms the underlying field to data space. For images it is the point-spread function (PSF), typically blurring the image. In the case of large-scale structure it is a dynamical transformation from an underlying density/potential field to the appropriate data space.

---

[1] The term image restoration is usually reserved for the case in which the image and data spaces are identical, while image reconstruction refers to other cases, for example, the data might be in the form of 1–d scans, while the image is 2–d. For our purposes the distinction is unimportant.

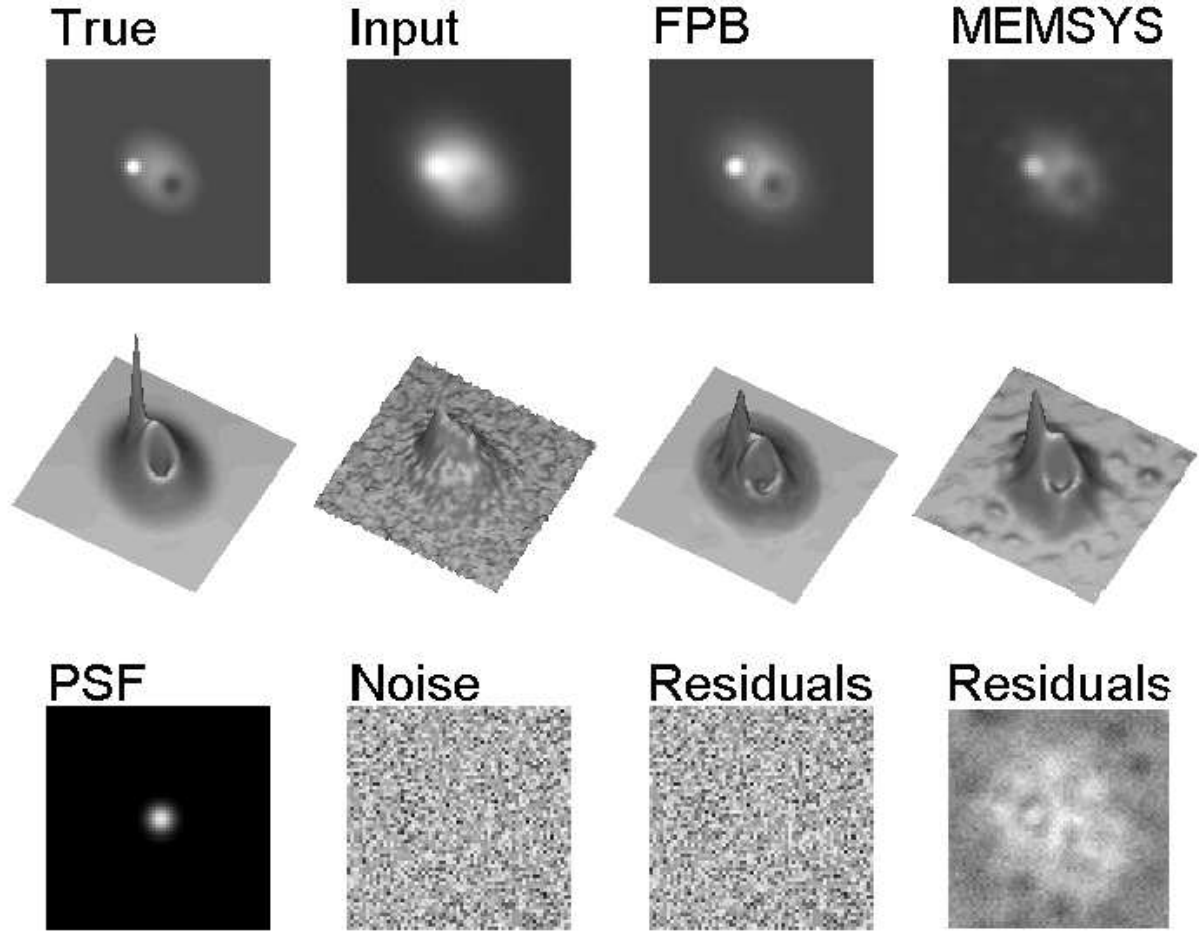

Figure 1: Comparison of pixon (FPB) and maximum entropy (MEMSYS) image restorations of a synthetic dataset, showing the true image, the noisy blurred input image, and the restored images and their residuals.

In solving Eq. (1), the noise is taken to be sufficiently well understood to allow the definition of a goodness-of-fit (GOF) criterion, such as $\chi^2$ or likelihood. It is further assumed that the functional form of the operator $H$ is known, although it may have unknown parameters which are to be determined from the data. In the cosmological case, in fact, the whole point is to determine some of the parameters of $H$, especially the cosmological density parameter, $\Omega$.

The purpose of this paper is to introduce some recent developments in image processing into the field of large-scale structure. §2 briefly presents some results of the new pixon method of Puetter and Piña [17, 16]. Since this method is based on concepts in information theory which are not generally known in the astronomical community, the paper then veers to a discussion of algorithmic information content, §3, complex adaptive systems, §4, and denoising using localized orthogonal functions such as wavelets and wave packets, §5. (For references see those sections.) A detailed presentation of the pixon method follows in §6, with application to large-scale structure in §7.

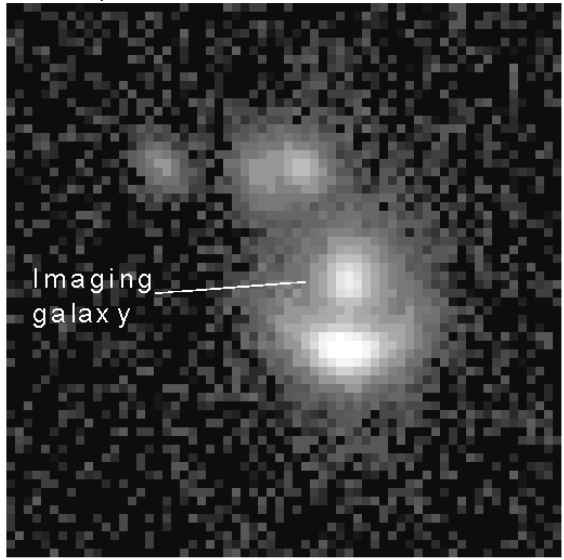
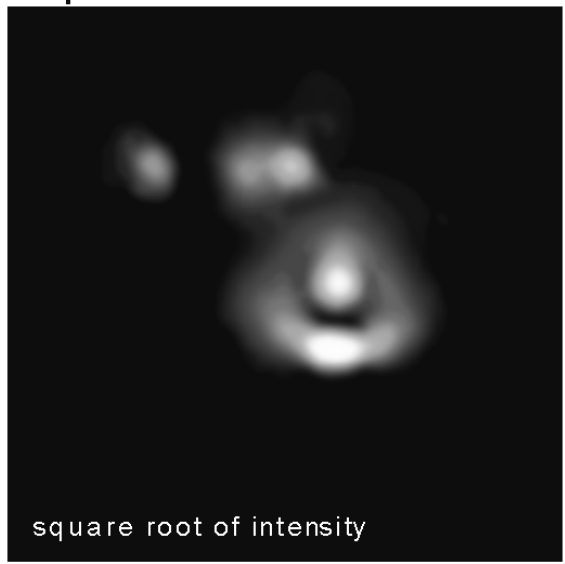
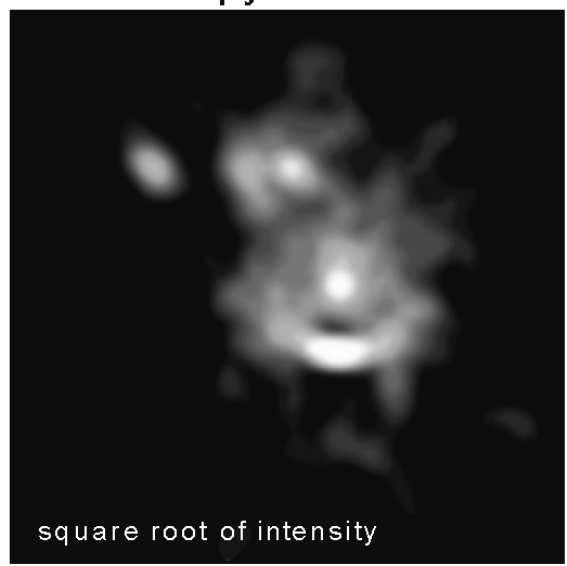

Figure 2: Reconstruction of the entire Einstein ring in FSC10214+4724, showing the raw $2\mu$m data taken at the Keck telescope and the reconstructions using elliptical pixons and maximum entropy. Note that the separation between the peak of the Einstein ring and the lensing galaxy in this ground-based observation is only $1''\!.5$.

## 2  Image reconstruction/restoration

When the data have high signal-to-noise ratios, Eq. (1) can be solved to obtain information about the image on scales smaller than the width of the PSF, provided that the data are sufficiently finely sampled. This inversion problem is a generalized deconvolution, and suffers from all its problems: (1) There may be less data points than image points, so the inversion is not unique. (2) The PSF is a smoothing operator, but noise introduces a high frequency component to the data. A straight inversion, even if unique, amplifies the noise. (3) A parametric fit of the image is always possible, but there has to be a model with which to parameterize. (4) If it is not known where to stop a parametric expansion, e.g., the order of a polynomial, a GOF criterion may be used, such as $\chi^2$ or likelihood, but the parameters have to be prioritized.

There is an extensive literature on inversion techniques, the state-of-the-art generally being taken to be the maximum entropy method [1, 15, 21, 13]. Recently Puetter and Piña introduced a new method using a Fractal Pixon Basis [17], to be explained in detail in §6, which is superior at detecting faint sources and resolving extended ones and far more robust against spurious artifacts and signal-correlated residuals. For a recent review with extensive examples see [16]. A synthetic example from that paper is reproduced in Fig. 1, comparing the pixon method with maximum entropy. A second comparison can be seen in the latest pixon restoration [18] of the full Einstein ring in the gravitational lens FSC10214+4724 [12], shown in Fig. 2.

## 3 Algorithmic information content

The algorithmic information content (AIC) of a dataset is defined as the length of the shortest program that prints the nonrandom content of the data and stops [22, 14, 2]. For example, consider a book whose first ten pages contain only the letter A, the next 10 pages B, and so on, ending with Z on page 260. The rather meager information in this book can be expressed very compactly by a program, making it unnecessary to present the entire data, i.e., the full book. In fact, the second sentence in this paragraph, although written in plain English instead of some computer language, is an accurate instruction for constructing the book.

This definition of information differs significantly from the original entropy introduced by Shannon [20] which uses fixed quanta of information, e.g., letters in the above example, or pixels in an image. Shannon's entropy is invariant under a random scrambling of the units of information and is insensitive to any *correlation* between them. It assigns the same entropy to the above-mentioned book and to one containing a completely random distribution of its letters. The AIC, on the other hand, seeks all possible regularities in order to minimize the description of the information; it recognizes that the book has some information in it beyond the fact that each letter occurs the same number of times. The maximum entropy method uses Shannon's definition of entropy and is therefore invariant under a scrambling of the pixels. As a result, it is not as effective in identifying extended structures.

Unlike the above example, real data always include some unwanted random noise, which is not part of the information content and is due to a physical limit (e.g., photon statistics) or instrumental noise. An artificially high noise level may even be set by the experimenter or analyzer to obscure uninteresting details that could be detected. In any event, the noise is characterized by a lack of correlation with the signal (information), and unpredictability, except in a statistical sense. (Systematic errors are not considered here.)

Clearly, failure to recognize the random component in a dataset leads to an erroneous determination of its information content. Consider, for example, a simple series in an IQ test: 1, 2, 3, 4, 5, 6, 7, 8, ?. The information content in this series can be identified by any child, who recognizes instantly that the next term in the series is 9. Now add noise, for example the following binary series was obtained by flipping a coin: 1, 1, 0, 1, 0, 1, 0, 0, ?. The real data are a sum of the true information and the noise, viz: 2, 3, 3, 5, 5, 7, 7, 8, ?, and the next term in the series is 9 or 10 with equal probability. Most people presented with this series, conditioned as they are on IQ tests, would attempt to find regularity in it assuming it to be noiseless—quite a challenge, actually. On the other hand, if the series were plotted against index, most scientists would recognize the noise component and correctly draw a regression line.

The point is that, even for such a simple series, and certainly for more complicated data, it is easy to mistake noise for signal and misinterpret the information. Furthermore, the tendency to make this mistake may be a function of how the data are viewed. Any search for information content, especially an automated search, must avoid this pitfall and correctly separate out the noise, so the remaining signal is noise-free. The question is how.

## 4 Complex adaptive systems

A somewhat surprising result is that, in the presence of noise, the least complex, most economical, description of information, i.e., the AIC, is also the most *faithful*; the data are not overfitted. Intuitive arguments illustrate this manifestation of Occam's razor. (For an example of a rigorous analysis of denoising see [9].) Consider an experiment in which $y_i$ are measured as a function of $x_i$, and suppose further that the true relation between them is a polynomial

$$y_i = p(x_i) + \epsilon_i \quad , \tag{2}$$

where $\epsilon_i$ is an error component. A simple polynomial least-squares fit gives the best coefficients, and a GOF test tells at what order to truncate the polynomial. But suppose that a least-squares fit is made with a sum of exponentials. More terms are typically needed to fit the data with the same $\chi^2$, and the fit is awkward, with correlated residuals. The more economical polynomial fit is also the more faithful one.

But in the absence of a known parameterization, how is one to choose from an infinity of possibilities? Gödel's incompleteness theorem actually guarantees that the entire information content of the data can never be found. The meaning of information content is therefore limited *a priori* by choosing an external language with which to describe it, and this language, in turn, restricts the allowed parameterizations. The choice of language does make a difference. The maximum entropy method makes no allowance for correlations between pixels in an image and assumes that the most probable one is completely gray. The pixon scheme, by contrast, favors images with large correlations.

A language that emphasizes correlations is more complex because of all the possible groupings of pixels. Moreover, these groupings are not known *a priori* and must be learned from the data. This is known as a complex adaptive system. For example, if a photograph of the mountains is taken, with perhaps some friends in the foreground, the information content might be the background mountains and valleys and foreground people. This says that the image is supposed to be made up of these objects, but not where in the image there are mountains and where valleys, and which people are located where in the foreground. The language can impose restrictions on the type of image, but in and of itself it does not specify all the details, which can only be learned from the data.

## 5 Localized orthogonal basis functions

For data whose expected values are functions of known variables, such as images as functions of position, the language of information has long been spectral. The data are described by means of a sum over spectral functions

$$y_i = \sum_k a_k f_k(x_i) + \epsilon_i \quad , \tag{3}$$

and the information content consists of the length of the series and its coefficients.

Many spectral functions are possible. For pulsed data, the Fourier spectrum might be all that is needed, while if spatial correlation is of no interest, the individual pixel intensities suffice. Images typically consist of extended contiguous structures which are both localized and have spatial correlations. The choice of spectral functions for this intermediate case is constrained by the uncertainty principle which determines the minimum volume of phase-space over which information can be defined. There has to be some give and take between correlation and localization of information.

In the last few years there has been great interest in wavelets [6] and wave packets [4, 5], spectral basis functions that are localized, orthogonal, and translation and dilation invariant. Their advantage is the speed of the transformations between the original and spectral domains, which require only $O(N \log_2 N)$ operations (like a fast Fourier transform). Denoising is achieved by transforming the noisy data into the spectral domain, applying either hard [10] or soft [8] thresholding to the resulting coefficients, thereby suppressing those smaller than a certain amplitude, and then transforming back into the original domain. These methods result additionally in data compression, since fewer spectral coefficients than data points need be kept. They have also been used to identify the important and noisy parts of initial conditions in dynamical problems, e.g., [11].

Despite their strong advantages, however, such denoising techniques have their limitations. First, the speed of the transformations depends on even sampling. More importantly, owing to their orthogonality, they must be both negative and positive. But images and densities are only positive, so a larger number of basis functions is required to "interfere away" the negative holes.[2] The orthogonal functions therefore form a less suitable basis in the AIC sense. This is borne out by the resulting denoised images, which continue to exhibit "visual artifacts" requiring special treatment [3].

## 6  Pixons

A representation of images which is closer to the AIC, and hence less likely to show residual artifacts, can be obtained by restricting the spectral functions to be positive. This approach, originated by Puetter and Piña [17], indeed leads to significant improvement in image reconstruction and restoration, as evidenced by several examples given in [16] and the image of the gravitational lens FSC10214+4724 [18] shown in Fig. 2. Fainter sources are detected, the angular resolution is better, and spurious features or signal-correlated residuals are largely absent.

The pixon method starts with a finely sampled positive pseudoimage, $\phi$, and smoothes it with a positive kernel function, $K$, to obtain the image.

$$I = K \otimes \phi \quad \Rightarrow \quad D = H \otimes K \otimes \phi + N \quad . \tag{4}$$

In its crudest form, the image model is made up of $P$ contiguous and nonoverlapping cells, called pixons, with constant intensity within each pixon. Unless these pixons have very complex structure, the AIC—the length of the shortest program required to specify the nonrandom part of the data—is likely to depend mainly on the total number of pixons, $P$, and be a monotonically rising function of it. The goal, then, is to minimize $P$, but without allowing the residuals to exceed the level expected for random noise. A straightforward way to achieve this is to minimize

$$\mathcal{L} = \text{GOF} + \lambda P \quad , \tag{5}$$

---

[2]Even for the velocity field, which can have both signs, there are constraints, for example on $\nabla \cdot \mathbf{v}$.

where GOF is a goodness-of-fit criterion, such as $\chi^2$ or likelihood, and $\lambda$ is a Lagrange multiplier adjusted to give a "reasonable" GOF. (It does not matter if $\lambda$ is attached to the GOF term or to P.) The procedure is completely analogous to the maximum entropy method; the difference is only in the choice of the entropy term, which in the pixon method seeks images with as much smoothness as possible.

A more refined approach limits the shapes that the kernel function can take, but allows its edges to be fuzzy, overlapping adjacent kernels. For example, the pixons used in the restoration of the gravitational lens FSC10214+4724, Fig. 2, were written as elliptical kernels

$$K(\mathbf{x}|\mathbf{y}) = \det\left(M(\mathbf{y})\right)^{1/2} f\big((\mathbf{x}-\mathbf{y}) \cdot M(\mathbf{y}) \cdot (\mathbf{x}-\mathbf{y})\big) \quad , \tag{6}$$

where $\mathbf{y}$ and $\mathbf{x}$ are the positions in pseudoimage and image spaces, respectively, $M(\mathbf{y})$ is a positive-definite, symmetric, matrix which determines the directions and lengths of the axes of the ellipse, and $f$ is the kernel shape function.

The definition of pixons is here a little more subtle, because of their fuzzy edges, but not very difficult. Note first that $P$ need only be specified to within an unknown constant of proportionality, since the latter can always be absorbed by the Lagrange multiplier, $\lambda$. A reasonable definition of a pixon density is therefore

$$p(\mathbf{y}) = \det\left(M(\mathbf{y})\right)^{1/2} \quad , \tag{7}$$

with the total number of pixons (the entropy term) being the integral

$$P = \int d\mathbf{y}\, p(\mathbf{y}) \quad . \tag{8}$$

The spectacular improvement brought by the pixon method to reconstructed image comes with a price tag of significant complication in numerical computations: (1) There are no simple orthogonal transformations between the original and spectral domains as for wavelets and wave packets. (2) In fact, since the kernel varies over the image, standard techniques for convolution do not apply either. (3) The minimization of $\mathcal{L}$ is nonlinear, with the number of variables proportional to the number of pixels. (4) Unless the signal-to-noise ratio is high, the minimization landscape is very complex, and a program may have to "slalom its way" in complicated terrain toward the minimal value.

When the signal-to-noise ratio (SNR) is not too low, a pixon reconstruction proceeds reasonably well by alternately minimizing the GOF term and $P$ [16]. More elaborate methods are needed for lower SNR. In the case of a $\gamma$-ray experiment with lower SNR [7], the derivation of the pixon map required mean field annealing [19], at yet higher computational cost. Effort is now underway by Puetter and the author to develop faster, and more general techniques for minimizing $\mathcal{L}$. Some progress has already been made, but the ultimate goal, a "black box" program which does not require fine adjustments by the user, is not yet at hand.

## 7  Application to large-scale structure

Galaxies in density and velocity surveys are unevenly sampled in space, both because of real density variations, and for practical reasons, particularly the increase of observational difficulty with distance. Trying to smooth the fields with a fixed window therefore requires compromises, which invariably lead to oversmoothing in some locations and undersmoothing in others.

For some applications, homogeneous smoothing is critical, for example in order to compare with a theoretical probability density function computed for a fixed smoothing window. In other cases, however, the ideal of homogeneous smoothing may be relaxed in favor of the maximum information contained in the data, introducing at each location the maximum smoothing permitted by the the local data.

Smoothing proceeds in analogy with image reconstruction, except that there is no blurring due to a point-spread-function. Instead, the operator $H$ takes the form of a dynamical transformation from an underlying density/potential field to the appropriate data space. For density determinations, the data are positions in redshift space, while for velocity surveys they are magnitudes and velocity widths. The pseudoimage is the density/potential defined on a fine grid, and the pixon kernel $K$ smoothes it to the maximum degree permitted by a GOF criterion for the data, typically a maximum likelihood estimator for density and a $\chi^2$ for velocity.

The result is a density/potential map with variable smoothing, which can only be interpreted together with its smoothing map. In particular, the comparison of two different density/potential maps is only meaningful when account is taken of differences in their smoothing maps. At first sight, this might seem to limit the usefulness of the method, particularly for the all-important comparison of density and velocity fields, used to determine the cosmological density parameter, $\Omega$. However, in a joint mapping, the same underlying density/potential field can be applied to both datasets. In a region in which the velocity data are more sparse, say, the smoothing scale is determined by the density, and the velocity data are insensitive to the fine structure, and vice versa where velocity data are more abundant. The quality of the fit can then be tested *a posteriori* by verifying that both datasets separately satisfy GOF criteria, and that the residuals are random, showing neither autocorrelation nor correlation with signal.

The success of the pixon method in image reconstruction suggests that, if all goes well, a cosmological pixon fit, which makes optimal use of all the data, could provide the most sensitive determination of $\Omega$ from redshift and velocity surveys.

**Acknowledgements.** I am indebted to R. Puetter for first introducing me to pixons and then for extensive discussions on the subject. This work was supported in part by NASA.